\documentclass[11pt,letterpaper]{article}

\usepackage{latexsym,multirow}
\usepackage{amssymb,amsmath, bm}
\usepackage{graphicx}

\usepackage{enumerate}

\usepackage{natbib}
\usepackage[pdftex, bookmarksopen=true, bookmarksnumbered=true,
pdfstartview=FitH, breaklinks=true, urlbordercolor={0 1 0}, citebordercolor={0 0 1}]{hyperref}
\usepackage{colortbl}
\usepackage{subfigure}
\usepackage[color,all,import,arrow]{xy}

\usepackage{dcolumn}
\newcolumntype{.}{D{.}{.}{-1}}
\newcolumntype{d}[1]{D{.}{.}{#1}}
\usepackage{theorem}
\theoremstyle{plain}
\theoremheaderfont{\scshape}
\newtheorem{assumption}{Assumption}
\newtheorem{example}{Example}
\newtheorem{definition}{Definition}
\newtheorem{corollary}{Corollary}

\newtheorem{theorem}{Theorem}

\newcommand{\ind}{\mbox{$\perp\!\!\!\perp$}}
\newcommand{\nind}{\mbox{$\not\perp\!\!\!\perp$}}

\usepackage{rotating}

\usepackage{arydshln}

\usepackage[compact]{titlesec}

\allowdisplaybreaks

\newcommand\spacingset[1]{\renewcommand{\baselinestretch}%
{#1}\small\normalsize}

\newcommand{\blind}{0}

\newcommand*{\QEDB}{\hfill\ensuremath{\square}}

\newcommand{\bone}{\mathbf{1}}
\newcommand{\red}{\color{red}}

\newcommand{\pr}{\text{pr}}

\newcommand{\E}{\mathbb{E}}
\newcommand{\bX}{\mathbf{X}}

\newcommand{\bV}{\bm{V}}
\newcommand{\bv}{\bm{v}}

\newcommand{\bW}{\mathbf{W}}

\begin{document} 

\newcommand{\tit}{Principal Fairness for Human and Algorithmic Decision-Making}
%
%
\spacingset{1.25}

\if0\blind

{\title{\bf\tit\thanks{We thank Hao Chen, Shizhe Chen, Christina
      Davis, Cynthia Dwork, Peng Ding, Robin Gong, Jim Greiner, Sharad
      Goel, Gary King, Jamie Robins, and Pragya Sur for comments and
      discussions.  We also thank anonymous reviewers of the Alexander
      and Diviya Magaro Peer Pre-Review Program at IQSS for valuable
      feedback.}}

\author{Kosuke
    Imai\thanks{Professor, Department of Government and Department of
      Statistics, Harvard University.  1737 Cambridge Street,
      Institute for Quantitative Social Science, Cambridge MA 02138.
      Email: \href{mailto:imai@harvard.edu}{imai@harvard.edu} URL:
      \href{https://imai.fas.harvard.edu}{https://imai.fas.harvard.edu}}
    \and Zhichao Jiang\thanks{Assistant Professor, Department of
      Biostatistics and Epidemiology, University of Massachusetts,
      Amherst MA 01003.}  
}

\date{
  First Draft: May 11, 2020 \\
  This Draft: \today
}

\maketitle

}\fi

\if1\blind
\title{\bf \tit}

\maketitle
\fi

\pdfbookmark[1]{Title Page}{Title Page}

\thispagestyle{empty}
\setcounter{page}{0}
         
\begin{abstract}
 Using the concept of principal stratification from the causal
 inference literature, we introduce a new notion of fairness, called
 principal fairness, for human and algorithmic decision-making. The
 key idea is that one should not discriminate among individuals who
 would be similarly affected by the decision. Unlike the existing
 statistical definitions of fairness, principal fairness explicitly
 accounts for the fact that individuals can be impacted by the
 decision. Furthermore, we explain how principal fairness differs
 from the existing causality-based fairness criteria. In contrast to
 the counterfactual fairness criteria, for example, principal
 fairness considers the effects of decision in question rather than
 those of protected attributes of interest. We briefly discuss how
 to approach empirical evaluation and policy learning problems under
 the proposed principal fairness criterion.

\end{abstract}

\newpage
\spacingset{1.5}

Although the notion of fairness has long been studied, the increasing
reliance on algorithmic decision-making in today's society has led to
the fast-growing literature on algorithmic fairness \citep[see
e.g.,][and references
therein]{corb:goel:18,baro:hard:nara:19,chou:roth:20,berk2021fairness,mitc:etal:21}.
In this paper, we introduce a new definition of fairness, called {\it
 principal fairness}, for human and algorithmic decision-making.
Unlike the existing {\it statistical fairness} criteria
\citep{hard:pric:sreb:16,chou:17,zafa:etal:17,john:lum:19}, principal
fairness incorporates causality into fairness. Furthermore, we explain
how principal fairness differs from the existing causality-based
fairness criteria. In particular, principal fairness differs from the
{\it counterfactual equalized odds} criteria in that it considers
joint potential outcomes and thus takes into account how the decision
affects the outcome \citep{coston2020counterfactual}. Moreover, in
contrast to the {\it counterfactual fairness} criteria
\citep{kusn:etal:17,nabi2018fair,zhan:bare:18,chiappa2019path},
principal fairness considers the effects of decision in question
rather than those of protected attributes of interest. We explore the
formal relations between principal fairness and these other fairness
criteria.

The key idea of principal fairness is that one should not discriminate
among individuals who would be similarly affected by the decision.
Consider a judge who decides, at a first appearance hearing, whether
to detain or release an arrestee pending disposition of any criminal
charges \citep[see][for a related empirical
study]{imai:etal:22}. Suppose that the outcome of interest is whether
the arrestee commits a new crime before the case is
resolved. According to principal fairness, the judge should not
discriminate between arrestees if they would behave in the same way
under each of two potential scenarios --- detained or released. For
example, if both of them would not commit a new crime regardless of
the decision, then the judge should not treat them
differently. Therefore, principal fairness is related to individual
fairness \citep{dwor:etal:12}, which demands that similar individuals
should be treated similarly. The critical difference is that for
principal fairness the similarity is measured based on the potential
(both factual and counterfactual) outcomes rather than observed
variables such as observed outcome, covariates, or any function of
them.

\section{Principal fairness}
\label{sec::decision}

We begin by formally defining principal fairness. Let
$D_i \in \{0, 1\}$ be the binary decision variable and
$Y_i \in \{0, 1\}$ be the binary outcome variable of interest. For
the simplicity of exposition, we assume that the outcome and treatment
variables are both binary, but as shown later, the framework can be
extended to other variable types. Following the standard causal
inference literature
\citep[e.g.,][]{neym:23,fish:35,rubi:74a,holl:86}, we use $Y_i(d)$ to
denote the potential value of the outcome that would be realized if
the decision is $D_i = d$ for $d=0,1$. Then, the observed outcome can
be written as $Y_i = Y_i(D_i)$.

Principal strata are defined as the joint potential outcome values,
i.e., $R_i = (Y_i(1), Y_i(0))$, \citep{fran:rubi:02}. Since any causal
effect can be written as a function of potential outcomes, e.g.,
$Y_i(1)-Y_i(0)$ and $Y_i(1)/Y_i(0)$, each principal stratum represents
how an individual would be affected by the decision with respect to
the outcome of interest. In other words, the principal strata contain
all the information about how the decision impacts the outcome. Unlike
the observed outcome $Y_i$, however, the potential outcomes, and hence
principal strata, represent the pre-determined characteristics of
individuals and are not affected by the decision. Moreover, since we
only observe one potential outcome for any individual, principal
strata are not directly observable.

In the criminal justice example, the principal strata are defined by
whether or not each arrestee commits a new crime under each of the two
scenarios --- detained or released --- determined by the judge's
decision. Let $D_i = 1$ ($D_i = 0$) represent the judge's decision to
detain (release) an arrestee, and $Y_i = 1$ ($Y_i = 0$) denote that
the arrestee commits (does not commit) a new crime. Then, the stratum
$R_i=(0, 1)$ represents the ``preventable'' group of arrestees who
would commit a new crime only when released, whereas the stratum
$R_i = (1, 1)$ is the ``dangerous'' group of individuals who would
commit a new crime regardless of the judge's decision. Similarly, we
may refer to the stratum $R_i=(0,0)$ as the ``safe'' group of
arrestees who would never commit a new crime, whereas the stratum
$R_i =(1,0)$ represents the ``backlash'' group of individuals who
would commit a new crime only when detained.\footnote{One could assume
 that an arrestee can never commit a new crime when detained,
 implying the absence of the backlash and dangerous groups. Here, we avoid such an assumption for
 the sake of generality. In an empirical application, we also find
 that a new crime can be committed even when an arrestee is detained
 \citep{imai:etal:22}.}

Principal fairness implies that the decision is independent of the
protected attribute within each principal stratum. In other words, a
fair decision-maker can consider a protected attribute only so far as
it relates to potential outcomes. We now give the formal definition
of principal fairness.
\begin{definition}[Principal fairness] \spacingset{1}
 \label{def::principal-decision} A decision-making
 mechanism satisfies principal fairness with respect to the outcome
 of interest and the protected attribute $A_i$ if the resulting
 decision $D_i$ is conditionally independent of $A_i$ within each
 principal stratum $R_i$, i.e.,
 $\Pr(D_i \mid R_i ,A_i) = \Pr(D_i \mid R_i).$
\end{definition}
Note that principal fairness requires one to specify the outcome of
interest as well as the attribute to be protected. As such, a
decision-making mechanism that is fair with respect to one outcome may
not be fair with respect to another outcome. Moreover, this
definition is generalizable to any treatment and outcome variable
types. For example, if the treatment is a continuous variable, there
exist an infinite number of principal strata, but the conditional
independence relation in Definition~\ref{def::principal-decision} is
still well defined.

\begin{table}[t!]
\begin{center}\spacingset{1}
\begin{tabular}{cccc}
\hline
 & & \multicolumn{2}{c}{Group A} \\
 & & $Y_i(0)=1 $ &  $Y_i(0)=0$ \\ \hline
& & Dangerous & Backlash \\
\multirow{2}{*}{$Y_i(1)=1$} & Detained ($D_i = 1$) & 120 & 30 \\
& Released ($D_i = 0$) &  30  &  30 \\ \cdashline{2-4}
& & Preventable & Safe \\
\multirow{2}{*}{$Y_i(1)=0$} & Detained ($D_i = 1$) & 70 & 30 \\  
         & Released ($D_i = 0$) &  70   & 120 \\ \hline 
 & & \multicolumn{2}{c}{Group B} \\
 & & $Y_i(0)=1 $ &  $Y_i(0)=0$ \\ \hline
& & Dangerous & Backlash \\
\multirow{2}{*}{$Y_i(1)=1$} & Detained ($D_i = 1$) & 80 & 20 \\
& Released ($D_i = 0$) &  20  & 20   \\ \cdashline{2-4}
& & Preventable & Safe \\
\multirow{2}{*}{$Y_i(1)=0$} & Detained ($D_i = 1$) & 80 & 40 \\  
         & Released ($D_i = 0$) &  80   &  160  \\ \hline          
\end{tabular}
\caption{Numerical illustration of principal fairness. Each cell
 represents a principal stratum defined by the values of two
 potential outcomes $(Y_i(1), Y_i(0))$, while two numbers within a
 cell represent the number of individuals detained $(D_i=1)$ and that
 of those released $(D_i=0)$, respectively. This example illustrates
 principal fairness because Groups~A~and~B have the same detention
 rate within each principal stratum.}
\label{tab:pf-ex}
\end{center}
\end{table}%

Table~\ref{tab:pf-ex} presents a numerical illustration, in which the
detention rate is identical between Groups~A~and~B within each
principal stratum. For example, within the ``dangerous'' stratum, the
detention rate is 80\% for both groups, while it is only 20\% for them
within the ``safe'' stratum. Indeed, the decision is independent of
group membership given principal strata, thereby satisfying principal
fairness.

\section{Comparison with the statistical fairness criteria}

In this section, we compare principal fairness with the existing
definitions of statistical fairness. 

\subsection{Three existing statistical fairness criteria}

We consider the following popular statistical fairness criteria.
\begin{definition}[Statistical Fairness] \label{def::stat-decision} \spacingset{1}
A decision-making mechanism is fair with respect to the outcome of
interest $Y_i$ and the protected attribute $A_i$ if the resulting
decision $D_i$ satisfies a certain conditional independence
relationship. Prominent examples of such relationships used in the
literature are given below. 
\begin{enumerate}
\item[(a)] {\sc Overall parity:} $\Pr( D_i \mid A_i) = \Pr(D_i)$
\item[(b)] {\sc Calibration:} $\Pr(Y_i \mid D_i, A_i) = \Pr(Y_i \mid D_i)$
\item[(c)] {\sc Accuracy:} $\Pr(D_i \mid Y_i, A_i) = \Pr(D_i \mid Y_i)$ 
\end{enumerate}
\end{definition}
In our example, suppose that the protected attribute is race. Then,
the overall parity implies that a judge should detain the same
proportion of arrestees across racial groups. In contrast, the
calibration criterion requires a judge to make decisions such that the
fraction of detained (released) arrestees who commit a new crime is
identical across racial groups. Finally, according to the accuracy
criterion, a judge must make decisions such that among those who
committed (did not commit) a new crime, the same proportion of
arrestees had been detained across racial groups.

\begin{table}[t!]
\begin{center} \spacingset{1}
\begin{tabular}{ccccc}
\hline
 & \multicolumn{2}{c}{Group A} & \multicolumn{2}{c}{Group B}\\
     & Detained & Released &  Detained &  Released \\ \hline
$Y_i=1$  & 150    & 100   &  100 &  100  \\ 
$Y_i=0$  & 100 & 150 & 120    &  180  \\ \hline    
\end{tabular}
\caption{Observed data calculated from Table~\ref{tab:pf-ex}. None of
 the statistical fairness criteria given in
 Definition~\ref{def::stat-decision} is met.} \label{tab:pf-obs}
\end{center}
\end{table}%

Principal fairness differs from these statistical fairness criteria in
that it accounts for how the decision affects the outcome. In
particular, although the accuracy criterion resembles principal
fairness, the former conditions upon the observed rather than
potential outcomes. Table~\ref{tab:pf-obs} presents the observed data
consistent with the numerical example shown in
Table~\ref{tab:pf-ex}. Although this example satisfies principal
fairness, it fails to meet any of the three statistical fairness
criteria. For example, among those who committed a new crime, the
detention rate is much higher for Group~A than Group~B. The reason is
that among these arrestees, the proportion of ``dangerous" individuals
is greater for Group~A than that for Group~B, and the judge is on
average more likely to issue the detention decision for these
individuals.

\subsection{Relationship between principal fairness and statistical
 fairness criteria}

How should we reconcile this tension between principal fairness and
the existing statistical fairness criteria? The tradeoffs between
different fairness criteria have been considered before in the
literature. As shown in the literature
\citep[e.g.,][]{chou:17,klei:mull:ragh:17, baro:hard:nara:19}, it is
generally impossible to simultaneously satisfy the three statistical
fairness criteria introduced in Definition~\ref{def::stat-decision}.
In some cases, however, principal fairness implies all three
statistical fairness criteria. The following theorem provides a
sufficient condition.
\begin{theorem}
 \label{thm::pf-sf} \spacingset{1}
 Suppose that $A_i \ind R_i$. Then, principal fairness in
 Definition~\ref{def::principal-decision} implies all three
 statistical definitions of fairness given in
 Definition~\ref{def::stat-decision}.
\end{theorem}

The condition states that different protected groups have the same
distribution of principal strata $R_i$. In the criminal justice
example, this means that no group is inherently more dangerous than
the other. This independence differs from the equal base rate
condition, i.e., $Y_i\ind A_i$, that has been identified in the
literature as a sufficient condition for simultaneously satisfying the
three statistical existing fairness criteria
\citep{klei:mull:ragh:17}. The equal base rate condition is based on
observed outcomes, which may be affected by the decision under
consideration. In contrast, our sufficient condition, $A_i \ind R_i$,
is about the independence between the protected attribute and
principal strata. Principal strata are based on potential outcomes,
which cannot be affected by the decision, and hence should be
considered as the characteristics of arrestees. As a result,
$A_i \ind R_i$ does not necessarily imply the equal base rate
condition, or vice versa. It can be shown, however, that if principal
fairness holds, $A_i \ind R_i$ also implies the equal base rate
condition. In other words, Theorem~\ref{thm::pf-sf} provides an
alternative condition under which statistical fairness criteria hold
simultaneously.

In many settings, it is reasonable to assume that the protected
attribute does not directly affect potential outcomes. In criminal
justice example, being a member of a particular racial group should not
make one inherently more dangerous. The protected attribute can,
however, affect potential outcomes through other mediating variables.
In particular, the existence of racial discrimination can yield an
association between race and various socio-economic variables, which
in turn generates the dependence between race and potential outcomes.
For this reason, the independence condition in
Theorem~\ref{thm::pf-sf} is likely to be violated in many
applications.


Thus, we further investigate the connection between principal fairness
and the statistical fairness criteria in more general settings without
requiring the independence condition $A_i \ind R_i$. Consider the
following monotonicity assumption.
\begin{assumption}[Monotonicity]
\label{asm::mon} \spacingset{1}
$$Y_i(1) \ \leq \ Y_i(0) \quad {\rm for\ all}\ i.$$
\end{assumption}
Assumption~\ref{asm::mon} is plausible in many applications when the
effect of the decision on the outcome is non-positive for all
individuals. In our criminal justice example, the assumption implies
that detention makes it no more likely for an arrestee to commit a new
crime in comparison to release. The following theorem establishes the
exact relationship between $\Pr(D_i \mid R_i,A_i)$ with
$\Pr(D_i, Y_i\mid A_i)$ under Assumption~\ref{asm::mon}.
\begin{theorem}
\label{thm::connection} \spacingset{1}
Under Assumption~\ref{asm::mon}, we have
\begin{eqnarray*}
\Pr(D_i=1 \mid R=(0,0),A_i) 
&=&1- \frac{\Pr(D_i=0, Y_i=0 \mid A_i)}{\Pr(R_i=(0,0) \mid A_i)},\\
\Pr(D_i=1 \mid R=(0,1),A_i) 
&=& \frac{\Pr(Y_i=0\mid A_i)-\Pr(R_i=(0,0)\mid A_i) }{\Pr(R_i=(0,1)\mid A_i) },\\
\Pr(D_i=1 \mid R=(1,1),A_i) 
&=&\frac{\Pr(D_i=1, Y_i=1 \mid A_i)}{\Pr(R_i=(1,1) \mid A_i)}.
\end{eqnarray*}
\end{theorem}
Proof is given in Appendix~\ref{app:equivalence}.
Theorem~\ref{thm::connection} shows that $\Pr(R_i\mid A_i)$ is the key
factor in connecting principal fairness to the statistical fairness
criteria. If $A_i$ is not independent of $R_i$, principal fairness and
the statistical fairness definitions do not imply each other. When
$A_i \ind R_i$ holds, however, principal fairness is equivalent to the
statistical fairness criteria under the monotonicity assumption. This
result is presented as the following corollary.
\begin{corollary} \spacingset{1}
\label{thm::decision-equivalence}
Suppose that $A_i\ind R_i$ holds. Then, under
Assumptions~\ref{asm::mon}, principal fairness is equivalent to the
three statistical fairness criteria given in
Definition~\ref{def::stat-decision}.
\end{corollary}

\section{Comparison with the existing causality-based fairness criteria}

We are not the first one to incorporate causality into the study of
algorithmic fairness. In this section, we explain how principal
fairness differs from the existing causality-based fairness criteria.

\subsection{Counterfactual equalized odds criterion}

The explicit conditioning of potential outcomes in fairness criteria
is not new. Independent of our work, \citet{coston2020counterfactual}
propose the following counterfactual equalized odds criterion,
\begin{equation}
 \Pr(D_i \mid Y_i(0) ,A_i) \ = \ \Pr(D_i \mid Y_i(0)).
\end{equation}
The authors justify conditioning on the potential outcome under the
control condition, $Y_i(0)$, by arguing that it represents a ``natural
baseline'' in most risk assessment settings.

Unlike the counterfactual equalized odds criterion, principal fairness
conditions on principal strata, which is defined by all potential
outcomes rather than a baseline potential outcome alone. This means
that in the case of a binary treatment, principal fairness includes
$Y_i(1)$ as well as $Y_i(0)$. The key idea is that principal fairness
considers how the decision impacts individuals, requiring the
comparison of all potential outcomes. In contrast, the counterfactual
equalized odds criterion focuses on the assessment of risk, which is
defined as the outcome in the absence of an intervention.

The difference between the two criteria can be illustrated via the
numerical example in Table~\ref{tab:pf-ex}. As explained earlier,
this example satisfies principal fairness, and yet it fails to meet
the counterfactual equalized odds criterion. For example, among those
who would commit a crime if released, the detention rate is higher for
Group A ($19/29$) than Group B ($16/26$). The reason is that those who
would commit a crime if released include both the ``dangerous'' and
``preventable'' individuals. The proportion of ``dangerous''
individuals is larger for Group A than that for Group B, and the judge
is more likely to impose a detention decision for these individuals.

The counterfactual equalized odds criterion could be viewed as a
special case of principal fairness when the decision is binary and the
potential outcome under the treatment condition $Y_i(1)$ is constant
across individuals. For example, if no individual can commit a new
crime when detained, the two criteria are equivalent. In our
empirical application, however, we find that a new crime can be
committed even when an arrestee is detained \citep{imai:etal:22}. In
addition, there are many settings where such an assumption is not
appropriate and one must consider how the decision affects different
individuals. They include the impacts of lending decisions on
household finance, and the effects of admissions decisions on future
wages.

In general, the following theorem establishes a sufficient condition
under which principal fairness implies the counterfactual equalized
odds criterion.
\begin{theorem} \spacingset{1}
 \label{thm::pf-codds}
Suppose that $Y_i(1)\ind A_i \mid Y_i(0)$. Then, principal fairness
implies $\Pr(D_i \mid Y_i(0) ,A_i) = \Pr(D_i \mid Y_i(0))$. 
\end{theorem}
Proof is given in Appendix~\ref{app:pf-codds}. This conditional
independence relation implies, in our example, that among those who
exhibit the same behavior under the release decision, the crime rate
under the detention decision is identical for Groups~A~and~B. This
condition is violated in many settings where the protected attribute
is associated with $Y_i(1)$ through variables other than $Y_i(0)$.

\subsection{Counterfactual fairness}
\label{sec:counterfactual}

In the algorithmic fairness literature, {\it counterfactual fairness}
represents one prominent fairness criterion that builds upon the
causal inference framework. \citet{kusn:etal:17} define the
counterfactual fairness by considering the potential decision when the
protected attributes are set to a fixed value. Under their definition,
a decision is counterfactually fair if a protected attribute does not
have a causal effect on the decision. In the criminal justice example,
counterfactual fairness implies that the decision an arrestee would
receive if he/she were white should be similar to the decision that
would be given if the arrestee were black. Formally, we can write this
criterion as,
\begin{equation*}
 \Pr\{D_i(a) = 1\} \ = \ \Pr\{D_i(a^\prime) = 1\}
\end{equation*}
for any $a \ne a^\prime$ where $D_i(a)$ represents the potential
decision when the protected attribute $A_i$ takes the value $a$.
Below, we briefly compare principal fairness with counterfactual
fairness.

First, while principal fairness is based on the statistical
independence between the {\it realized} decision $D_i$ and the
protected attribute $A_i$, counterfactual fairness requires the
distribution of {\it potential} decision to be equal across the values
of the protected attribute. Counterfactual fairness can be defined at
an individual level, i.e., $D_i(a) = D_i(a^\prime)$, which demands
that, for example, an arrestee should receive the same decision even
if he/she were to belong to a different racial group. In contrast,
principal fairness, like existing statistical fairness criteria, is
fundamentally a group-level notion and cannot be defined at an
individual level. Ensuring group-level fairness may not guarantee
individual-level fairness.
 
Second, while principal fairness considers the potential outcomes with
respect to different decisions, counterfactual fairness is based on
the potential outcomes regarding different values of protected
attribute. In the causal inference literature, some advocated the
mantra ``no causation without manipulation'' by pointing out the
difficulty of imagining a hypothetical intervention of altering one's
immutable characteristics such as race and gender
\citep[e.g.,][]{holl:86}. In addition, causal mediation analysis
relies upon the so-called ``cross-world'' independence assumption that
cannot be satisfied even when the randomization of mediators is
possible \citep{rich:robi:13}. Addressing these issues often requires
one to consider alternative causal quantities such as the causal
effects of perceived attributes \citep{grei:rubi:11} and stochastic
intervention of mediators \citep{jack:vand:18}. In contrast, principal
fairness avoids these conceptual and identifiability issues and can be
evaluated under the widely used unconfoundedness assumption.

Finally, recall that as shown in Theorem~\ref{thm::decision},
principal fairness implies all other statistical fairness criteria
under $A_i\ind R_i$. However, even under this assumption, principal
fairness neither implies nor is implied by counterfactual fairness. As
the following example illustrates, a decision rule that directly
depends on the protected attribute can satisfy principal fairness
while failing to meet counterfactual fairness. Alternatively, a
decision rule that does not depend on the protected attribute can meet
counterfactual fairness but may fail to meet principal fairness.

\begin{example} 
\label{ex1}
Consider a population characterized by the following
 distributions of principal strata
 $R \in \{(0,0),(1,0),(0,1), (1,1)\}$, the protected attribute
 $A \in \{0, 1\}$, and the covariate $X \sim \text{Unif}(0,1)$,
\begin{eqnarray*}
 \Pr(A = 1 \mid X) & = & X, \\ 
\Pr(R=(1,1)\mid A=a,X) &=& \Pr(R=(0,0)\mid A=a,X) \ = \ 0.3, \\
\Pr(R=(1,0)\mid A=a,X ) &=& \Pr(R=(1,0)\mid A=a,X ) \ = \ 0.2
\end{eqnarray*}
for $a=0,1$. This implies $A \ind R$.
Consider the decision rule of the following form,
$D= \bone\{\alpha X +\beta A \geq 1\}$. Suppose $\alpha = 5/2$ and
$\beta = -1$. Then, we have,
\begin{eqnarray*}
 \Pr\{D(1) = 1\} & = & 0.2,\quad \Pr\{D(0) = 1\} \ = \ 0.6, \\
 \Pr(D=1 \mid R=r,A=1) &=& \Pr( X\geq 0.8 \mid A=1)\ =\ 0.36,\\
 \Pr(D=1 \mid R=r,A=0) &=& \Pr( X\geq 0.4 \mid A=0)\ =\ 0.36. 
\end{eqnarray*}
Thus, the decision rule violates counterfactual fairness while
satisfying principal fairness. In contrast, consider $\alpha = 5/2$
and $\beta = 0$. Then, we have,
\begin{eqnarray*}
 \Pr\{D(1) = 1\} & = & \Pr\{D(0) = 1\} \ = \ 0.6, \\
 \Pr(D=1 \mid R=r,A=1) &=& \Pr( X\geq 0.4 \mid A=1)\ =\ 0.84,\\
 \Pr(D=1 \mid R=r,A=0) &=& \Pr( X\geq 0.4 \mid A=0)\ =\ 0.36.
\end{eqnarray*}
Thus, the decision rule violates principal fairness while satisfying
counterfactual fairness.

 
\end{example}

\section{Conditional fairness criteria}

Although we have so far focused on fairness criteria based on marginal
distributions, policy makers and researchers may be interested in
evaluating fairness within each subpopulation defined by a set of
pre-treatment covariates. The importance of such conditioning
covariates has been recognized in the algorithmic fairness literature.
Specifically, even when a statistical fairness criterion holds
conditional on a set of covariates, the same criterion may not be
satisfied without those conditioning covariates. The reason is that
these conditioning covariates may be correlated with the protected
attribute itself. This problem is called infra-marginality in the
literature and applies to all statistical fairness criteria including
principal fairness \citep{corb:goel:18}. The infra-marginality
problem simply reflects an unavoidable fact that conditional
independence does not necessarily imply marginal independence and vice
versa.

The following theorem shows that if the conditioning covariates
eliminate the dependence between the protected attribute and principal
stratum, then conditional on these covariates, principal fairness
implies all three statistical definitions of fairness and the
counterfactual equalized odds criterion.
\begin{theorem} \spacingset{1}
 \label{thm::decision}
 Suppose that there exist a set of variables $\bW_i$ such that
 $A_i \ind R_i \mid \bW_i$. Then, conditional on $\bW_i$, principal
 fairness implies the counterfactual equalized odds criterion and all
 three statistical definitions of fairness. That is,
 $\Pr(D_i \mid R_i, \bW_i, A_i) = \Pr(D_i \mid R_i, \bW_i)$ implies
 $\Pr(D_i \mid Y_i(0), \bW_i, A_i) = \Pr(D_i \mid Y_i(0), \bW_i)$,
 $\Pr(D_i \mid \bW_i, A_i) = \Pr(D_i \mid \bW_i)$,
 $\Pr(Y_i \mid D_i, \bW_i, A_i) = \Pr(Y_i \mid D_i, \bW_i)$, and
 $\Pr(D_i \mid Y_i, \bW_i, A_i) = \Pr(D_i \mid Y_i, \bW_i)$.
 Moreover, if Assumption~\ref{asm::mon} also holds, then principal
 fairness is equivalent to all three statistical definitions of
 fairness conditional on $\bW_i$.
\end{theorem}
Proof is given in Appendix~\ref{app:decision}. 

The conditional independence $A_i \ind R_i \mid \bW_i$ means that no
racial group is inherently more dangerous than other groups once we
account for relevant factors $\bW_i$. In a causal model, the absence
of direct effect of $A_i$ on $R_i$ implies the existence of $\bW_i$
that satisfies $A_i \ind R_i \mid \bW_i$ where $\bW_i$ can include
mediators as well as common causes. The lack of the direct effect of
race can be viewed as an axiomatic assumption that belonging to a
particular racial group does not make one inherently more dangerous
than members of other racial groups.

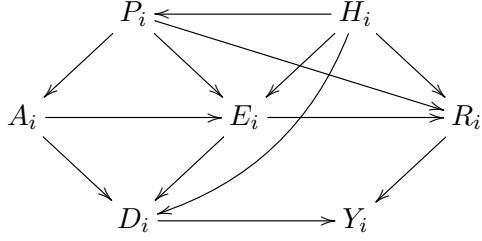
\begin{figure}[t]
\centering\spacingset{1}
$$
\xymatrix{
 & P_{i} \ar[ld]\ar[rd] \ar[rrrd] & & H_{i}\ar[ll] \ar@/^1.3pc/[lldd] \ar[ld] \ar[rd]& \\
A_i \ar[rr] \ar[rd]& & E_{i} \ar[ld]\ar[rr] && R_i \ar[ld]\\
  &D_i \ar[rr]& &Y_{i}&
}
$$
\caption{Direct acyclic graph for the relationship between the
 protected attribute $A_i$ and principal strata $R_i$. In the
 criminal justice application, $A_i$ represents the race of an
 arrestee, $R_i$ is their risk category (safe, preventable,
 dangerous, and backlash), $D_i$ represents the decision of judge,
 $P_{i}$ represents parents' characteristics including their
 attributes and socioeconomic status (SES), $E_{i}$ represents
 arrestee's own experiences such as SES, and $H_{i}$ represents
 historical processes. Finally, $Y_{i}$ is indicator of committing a
 new crime, which is a deterministic function of judge's decision
 $D_i$ and risk category $R_i$. The conditional independence $R_i \ind A_i \mid \bW_i$ holds with
 $\bW_i = (H_{i},P_{i},E_{i})$.}
\label{fig::asm1}
\end{figure}

For illustration, consider the causal model, shown as a directed
acyclic graph in Figure~\ref{fig::asm1}, in the context of the
criminal justice example. The race of an arrestee, $A_i$, is affected
by his/her parents' characteristics including their attributes and
social economic status (SES), $P_{i}$. The arrestee's own
experiences, $E_{i}$, are influenced by their race, $A_i$, their
parents' characteristics, $P_{i}$, and the historical processes such
as slavery and Jim Crow laws, $H_{i}$, which also affect the parents'
characteristics.

Under this causal model, all of these three covariates affect the risk
category of arrestee (principal strata; i.e., safe, preventable,
dangerous, and backlash), $R_i$, whereas the judge's decision, $D_i$,
is affected by the race, the experiences, and the historical
processes. The key assumption of the model is that the arrestee's race
does not {\it directly} affect their risk category, as indicated by
the absence of an arrow between these two variables. As a result,
under this model, the arrestee's race is conditionally independent of
risk category, i.e., $R_i \ind A_i \mid \bW_i$, where
$\bW_i = (H_i, P_{i}, E_{i})$. In other words, once we account for
these factors, no racial group has an innate tendency to be dangerous
relative to the other groups.

Theorem~\ref{thm::decision} shows that once we condition on $\bW_i$
that satisfies $A_i\ind R_i \mid \bW_i$, principal fairness implies
the counterfactual equalized odds criterion and all statistical
fairness criteria. However, this result should not be used to justify
the appropriateness of conditioning on $\bW_i$. The reason is that
the inclusion of conditioning covariates in fairness criteria can lead
to discrimination based on those variables. If the conditioning
covariates are good proxy variables for the protected attribute, then
any conditional fairness criteria could lead to discrimination against
those groups who should be protected. Thus, the choice of
conditioning covariates must be made with special care
\citep{kilbertus2017avoiding,beutel2019putting}.

Finally, the conditioning covariates also play an important role in
counterfactual fairness as well but for a different reason. Unlike
principal fairness or statistical fairness definitions, one cannot
simply condition on covariates that are affected by the protected
attribute because this would induce a post-treatment bias \citep[see
e.g.,][]{kilbertus2017avoiding, knox2019administrative}. To address
this issue, researchers have considered path-specific effects through
the framework of causal mediation analysis
\citep[e.g.,][]{nabi2018fair,zhan:bare:18,chiappa2019path}. In such an
analysis, a key question is which mediators should be included.

To further illustrate the difference between counterfactual fairness
and principal fairness with conditioning, we again consider the causal
model shown in Figure~\ref{fig::asm1}. Suppose we would like to
condition on $E_i$. Then counterfactual fairness requires that the
race has no effect on the decision other than through this
variable. Because the race can only affect the decision directly or
through $E_i$, counterfactual fairness is violated conditional on
$E_i$ due to the existence of the direct effect. In contrast,
principal fairness may still hold conditional on $E_i$ if the
association from the direct effect of $A_i$ on $D_i$ cancels out with
the association from the common cause $H_i$. Consistent with
Example~\ref{ex1}, a decision rule that directly depends on the
protected attribute can satisfy principal fairness.

\section{Empirical evaluation and policy learning under principal fairness}
\label{sec:evaluation}

Finally, we discuss how to use the above theoretical results in
empirical studies. We first show how to empirically assess the
independence conditions in
Theorems~\ref{thm::pf-codds}~and~\ref{thm::decision}, i.e.,
$Y_i(1)\ind A_i \mid Y_i(0)$ and $A_i \ind R_i$. To do this, we must
identify the distribution of the principal stratum within each group
defined by the protected attribute. We begin by introducing the
following unconfoundedness assumption, which is widely used in the
causal inference literature.
\begin{assumption}[Unconfoundedness]
\label{asm::ignorability} 
$Y_i(d) \ \ind \ D_i \mid \bX_i$ for any $d$.
\end{assumption}
Assumption~\ref{asm::ignorability} holds if $\bX_i$ contains all the
information used for decision-making. In practice, if we are unsure
about whether the protected attribute is used for decision-making, we
may still include it in $\bX_i$ to make the unconfoundedness
assumption more plausible \citep{vand:shpi:11}.  

The next theorem shows that under
Assumptions~\ref{asm::mon}~and~\ref{asm::ignorability}, the evaluation
of the independence relations, $Y_i(1)\ind A_i \mid Y_i(0)$ and
$A_i \ind R_i$, reduces to the estimation of conditional probability,
$\Pr(Y_i = 1 \mid D_i, \bX_i)$, from the observed data.
\begin{theorem}[Empirical Evaluation of $Y_i(1)\ind A_i \mid Y_i(0)$ and $A_i\ind R_i$]
\label{thm::indep} 
Under Assumptions~\ref{asm::mon}~and~\ref{asm::ignorability}, we have 
\begin{eqnarray*}
\Pr(Y_i(1)=1\mid A_i,Y_i(0)=1) &=& \frac{m_1(A_i)}{m_0(A_i)},\\
\Pr(R_i=(0,0)\mid A_i) &=& 1-m_0(A_i),\\
\Pr(R_i=(0,1)\mid A_i) &=&m_0(A_i)-m_1( A_i),\\
\Pr(R_i=(1,1)\mid A_i) &=& m_1(A_i).
\end{eqnarray*}
where $m_d(A_i) = \E\{\Pr( Y_i=1 \mid D_i=d, \bX_i)\mid A_i\}$.
\end{theorem}
Proof is given in Appendix~\ref{app:indep}. Theorem~\ref{thm::indep}
shows that we can empirically evaluate the validity of
$Y_i(1)\ind A_i \mid Y_i(0)$ and $A_i \ind R_i$ by checking whether
the distribution of principal strata $R_i$ depends on the protected
attribute $A$. The result also holds conditional on any covariates
that are included in $\bX_i$.

Second, we consider the empirical evaluation of principal fairness.
Combining Theorems~\ref{thm::connection}~and~\ref{thm::indep}, the
following corollary shows that the same assumptions used in
Theorem~\ref{thm::indep} are sufficient for identifying the
conditional distribution of decision $D_i$ given the principal strata
and the protected attribute. Using this conditional distribution, one
can empirically assess the principal fairness of the decision.
\begin{corollary}[Empirical Evaluation of Principal Fairness]
\label{cor::identification} 
Under Assumptions~\ref{asm::mon}~and~\ref{asm::ignorability}, we have 
\begin{eqnarray*}
 \Pr\{D_i=1 \mid R_i=(0, 0),A_i\} 
&=&1-\frac{\Pr(D_i=0, Y_i=0\mid A_i) }{ 1-m_0(A_i)},\\
 \Pr\{D_i=1 \mid R_i=(0, 1),A_i\} 
&=& \frac{m_0(A_i)-\Pr(Y_i=1\mid A_i) }{m_0(A_i)-m_1(A_i)},\\
\Pr\{D_i=1 \mid R_i=(1,1), A_i\}
&=&\frac{\Pr(D_i=1, Y_i=1\mid A_i) }{ m_1(A_i)}.
\end{eqnarray*}
\end{corollary}
The formulas also hold conditional on any covariates that are included
in $\bX_i$ and thus allow for the evaluation of conditional principal
fairness. 

Finally, we consider policy learning under principal fairness. For
simplicity, we focus on a deterministic policy $D_i = \delta(\bV_i)$,
where $\bV_i$ represents the covariates used for making
decisions. Suppose that the protected attribute is binary. Then,
principal fairness requires the decision rule $ \delta(\bV_i)$ to
satisfy the following equality constraint,
$\Pr( \delta(\bV_i) \mid R_i,A_i=1) = \Pr( \delta(\bV_i) \mid
R_i,A_i=0)$. This constraint may be difficult to satisfy due to the
fact that $R_i$ is an unobserved variable. The following theorem
expresses this probability, $\Pr( \delta(\bV_i) \mid R_i,A_i=1)$, in a
different form that only depends on observed variables.
\begin{theorem}
\label{thm::learning}
Suppose that Assumptions~\ref{asm::mon} holds and the decision is a
function of $\bV_i$, i.e., $D_i = \delta(\bV_i)$. Then, we have,
\begin{eqnarray*}
&&\Pr( \delta(\bV_i)=1 \mid R_i=r, A_i)\ =\  \E\left[ \frac{e_r(\bV_i,A_i)}{ \E\{e_r(\bV_i,A_i) \mid A_i \}} \delta(\bV_i) \mid A_i \right],
\end{eqnarray*}
for $r =(0, 0), (0, 1)$, and $(1, 1)$ where
$e_r(\bV_i,A_i) = \Pr(R_i=r \mid \bV_i,A_i)$.
\end{theorem}

The identification formulas for $e_r(W_i,A_i) $ are given in
Theorem~\ref{thm::indep}. When we know which covariates are used in
the decision $D_i$, these identification formulas provide an
alternative way to evaluate principal fairness in addition to
Corollary~\ref{cor::identification}. Specifically,
Theorem~\ref{thm::learning} shows that $\Pr( D_i \mid R_i=r, A_i)$ is
equal to the decision probability within each protected group in a
weighted population. The weights depend on the proportions of
principal strata given the covariates and the protected
attribute. Therefore, to learn a policy that satisfies principal
fairness, one could first estimate $e_r(\bV_i,A_i)$ using
Theorem~\ref{thm::indep} and then augment the existing fairness-aware
policy learning approaches with statistical fairness constraints based on the estimated weights
\citep[e.g.,][]{kamishima2011fairness,agarwal2018reductions,celis2019classification}.


\section{Concluding Remarks}

To assess the fairness of human and algorithmic decision-making, we
may wish to consider how the decisions themselves affect
individuals. This requires the notion of fairness to be placed in the
causal inference framework. In a separate work, we apply the idea of
principal fairness to the common settings, in which humans make
decisions partly based on algorithmic recommendations
\citep{imai:etal:22}. Since human decision-makers rather than
algorithms ultimately impact individuals, one must assess whether
algorithmic recommendations improve the fairness of human
decisions. We empirically examine this issue through the experimental
evaluation of the pre-trial risk assessment instrument widely used in
the US criminal justice system.

The difference between principal fairness and counterfactual equalized
odds criterion sheds light on the predictive performance evaluation of
algorithmic risk assessments. The current literature focuses on the
prediction accuracy of $Y_i(0)$ when evaluating algorithmic fairness
under the counterfactual equalized odds criterion
\citep[e.g.,][]{coston2020counterfactual,mitc:etal:21}. However, in
general, $Y_i(0)$ alone does not fully characterize counterfactual
outcomes: individuals with the same value of $Y_i(0)$ may differ in
the value of $Y_i(d)$ where $d \ne 0$. Principal fairness generalizes
counterfactual equalized odds criterion by considering principal
strata which depend on all potential outcomes. In particular, the
evaluation of algorithmic decision or recommendation requires one to
condition on principal strata rather than the observed outcome or a
single potential outcome.

Finally, although this paper focuses on the introduction of principal
fairness as a new fairness concept, much work remains to be done. In
particular, future work should consider the development of algorithms
that achieve principal fairness. Another possible direction is the
extension of principal fairness to a dynamic system. As pointed out
by \cite{chou:roth:20} and \cite{damo:etal:20}, real-world
algorithmic systems operate in complex environments that are
constantly changing, often due to the actions of algorithms
themselves. Therefore, an explicit consideration of the dynamic causal
interactions between algorithms and human decision-makers can help us
develop long-term fairness criteria.

%
%
%
%
%

\bigskip

\pdfbookmark[1]{References}{References}
\bibliographystyle{Chicago}
\bibliography{fairness-ref,../PSA/PSA-ref,my,imai}

\newpage

\appendix
\setcounter{page}{1}

\setcounter{equation}{0}
\setcounter{figure}{0}
\setcounter{theorem}{0}
\setcounter{lemma}{0}
\setcounter{section}{0}
\renewcommand {\theequation} {S\arabic{equation}}
\renewcommand {\thefigure} {S\arabic{figure}}
\renewcommand {\thetheorem} {S\arabic{theorem}}
\renewcommand {\thelemma} {S\arabic{lemma}}
\renewcommand {\thesection} {S\arabic{section}}

\begin{center}
  \LARGE {\bf Supplementary Appendix}
\end{center}
\spacingset{1.1}

\section{Proof of Theorem~\ref{thm::pf-codds}}
\label{app:pf-codds}
By the law of total probability, we have 
\begin{eqnarray*}
\Pr(D_i \mid Y_i(0),A_i) &=& \sum_{y_1=0,1}\Pr(D_i \mid Y_i(1)=y_1, Y_i(0),A_i) \Pr(Y_i(1)=y_1\mid Y_i(0),A_i)\\
&=&\sum_{y_1=0,1}\Pr(D_i \mid Y_i(1)=y_1, Y_i(0)) \Pr(Y_i(1)=y_1\mid Y_i(0))\\
&=&\Pr(D_i \mid Y_i(0)),
\end{eqnarray*}
where the second equality follows from principal fairness and $Y_i(1)\ind A_i \mid Y_i(0)$.  \QEDB

\section{Proof of Theorem~\ref{thm::pf-sf}}
\label{app:decision}
We prove a more general version of Theorem~\ref{thm::pf-sf} with any variables $\bV_i$ in the conditioning set. That is, 
under $A_i\ind R_i \mid \bV_i$, principal fairness in
implies all three
 statistical definitions of fairness conditional on $\bV_i$.

Because the observed stratum $(D_i=1, Y_i=1)$ is a mixture of principal strata $R_i=(1,0),(1,1)$, we have
\begin{eqnarray*}
&&\Pr(D_i=1, Y_i= 1 \mid \bV_i,A_i)\\
 &=& \Pr(D_i=1, R_i=(1,0) \mid \bV_i,A_i)+ \Pr(D_i=1, R_i=(1,1) \mid \bV_i,A_i)\\
& =&\Pr(D_i=1 \mid  R_i=(1,0), \bV_i,A_i)\Pr(R_i=(1,0) \mid \bV_i,A_i)\\
&&+\Pr(D_i=1 \mid  R_i=(1,1), \bV_i,A_i)\Pr(R_i=(1,1) \mid \bV_i,A_i)\\
& =&\Pr(D_i=1 \mid  R_i=(1,0), \bV_i)\Pr(R_i=(1,0) \mid \bV_i)\\
&&+\Pr(D_i=1 \mid  R_i=(1,1), \bV_i)\Pr(R_i=(1,1) \mid \bV_i)\\
 &=& \Pr(D_i=1, R_i=(1,0) \mid \bV_i)+ \Pr(D_i=1, R_i=(1,1) \mid \bV_i)\\
&=& \Pr(D_i=1, Y_i= 1 \mid \bV_i),
\end{eqnarray*}
where the third equality follows from principal fairness and the
assumption $A_i \ind R_i \mid \bV_i$.  Similarly, we can show
\begin{eqnarray}
\label{eqn::pDY}
 \Pr(D_i=d, Y_i =y\mid \bV_i,A_i)= \Pr(D_i=d, Y_i=y \mid \bV_i)
\end{eqnarray}
for $d,y=0,1$.
Therefore, we have 
\begin{eqnarray}
\nonumber \Pr(D_i \mid \bV_i,A_i) &=&  \Pr(D_i, Y_i=1\mid \bV_i,A_i) + \Pr(D_i, Y_i=0 \mid \bV_i,A_i) \\
  \nonumber&=&  \Pr(D_i, Y_i=1\mid \bV_i) + \Pr(D_i, Y_i=0 \mid \bV_i) \\
\label{eqn::pD} &=& \Pr(D_i \mid \bV_i),
\end{eqnarray}
and
\begin{eqnarray}
\nonumber \Pr(Y_i \mid \bV_i,A_i) &=&  \Pr(D_i=1, Y_i\mid \bV_i,A_i) + \Pr(D_i=0, Y_i \mid \bV_i,A_i) \\
  \nonumber&=&  \Pr(D_i=1, Y_i\mid \bV_i) + \Pr(D_i=0, Y_i \mid \bV_i) \\
\label{eqn::pY} &=& \Pr(Y_i \mid \bV_i).
\end{eqnarray}
Then, from Equations~\eqref{eqn::pDY}~and~\eqref{eqn::pD}, we have 
$ \Pr(Y_i \mid D_i, \bV_i,A_i)=  \Pr(Y_i \mid D_i, \bV_i)$, and 
from Equations~\eqref{eqn::pDY}~and~\eqref{eqn::pY}, we have 
$ \Pr(D_i \mid Y_i, \bV_i,A_i)=  \Pr(D_i \mid Y_i, \bV_i)$.  \QEDB
%
\section{Proof of Theorem~\ref{thm::connection}}
\label{app:equivalence}


We prove a more general version of  Theorem~\ref{thm::connection} with any variables $\bV_i$ in the conditioning set.
From Assumption~\ref{asm::mon}, we obtain
\begin{eqnarray*}
\Pr(D_i=1 \mid R_i=(0,0),\bV_i, A_i)&=&1- \frac{\Pr(D_i=0, R_i=(0,0) \mid \bV_i,A_i)}{\Pr(R_i=(0,0) \mid \bV_i,A_i)}=1- \frac{\Pr(D_i=0, Y_i=0 \mid \bV_i,A_i)}{\Pr(R_i=(0,0) \mid \bV_i,A_i)},  \\
\Pr(D_i=1 \mid R_i=(1,1),\bV_i, A_i)&=&\frac{\Pr(D_i=1, R_i=(1,1)\mid \bV_i,A_i)}{\Pr(R_i=(1,1) \mid \bV_i,A_i)}=\frac{\Pr(D_i=1, Y_i=1 \mid \bV_i,A_i)}{\Pr(R_i=(1,1) \mid \bV_i,A_i)},
\end{eqnarray*} 
and
\begin{eqnarray*}
&&\Pr(D_i=1 \mid R_i=(0,1),\bV_i, A_i)\\
&=&\frac{\Pr(D_i=1, R_i=(0,1) \mid \bV_i,A_i)}{\Pr(R_i=(0,1) \mid \bV_i,A_i)}\\
&=&\frac{\Pr(D_i=1 \mid \bV_i,A_i)-\Pr(D_i=1, R_i=(1,1) \mid \bV_i,A_i)} {\Pr(R_i=(0,1) \mid \bV_i,A_i)}-\frac{\Pr(D_i=1, R_i=(0,0) \mid \bV_i,A_i)} {\Pr(R_i=(0,1) \mid \bV_i,A_i)}\\
&=&\frac{\Pr(D_i=1 \mid \bV_i,A_i)-\Pr(D_i=1, Y_i=1 \mid \bV_i,A_i)} {\Pr(R_i=(0,1) \mid \bV_i,A_i)}\\
&&-\frac{\Pr(R_i=(0,0) \mid \bV_i,A_i)- \Pr(D_i=0, R_i=(0,0) \mid \bV_i,A_i)} {\Pr(R_i=(0,1) \mid \bV_i,A_i)}\\
&=&\frac{\Pr(D_i=1, Y_i=0\mid \bV_i,A_i)} {\Pr(R_i=(0,1) \mid \bV_i,A_i)}-\frac{\Pr(R_i=(0,0) \mid \bV_i,A_i)- \Pr(D_i=0, Y_i=0 \mid \bV_i,A_i)} {\Pr(R_i=(0,1) \mid \bV_i,A_i)}\\
&=&\frac{\Pr(Y_i=0\mid \bV_i,A_i)-\Pr(R_i=(0,0) \mid \bV_i,A_i)}{\Pr(R_i=1 \mid \bV_i,A_i)}.
\end{eqnarray*}
\QEDB

%

 \section{Proof of Theorem~\ref{thm::indep}}
 \label{app:indep}
Under Assumption~\ref{asm::mon}, we have
\begin{eqnarray}
\label{eqn::R00-W} \Pr(R_i=(0,0) \mid A_i)&=& \Pr(Y_i(0)=0 \mid A_i),\\
\label{eqn::R01-W}  \Pr(R_i=(0,1) \mid A_i)&=& \Pr(Y_i(0)=1 \mid A_i)- \Pr(Y_i(1)=1 \mid A_i),\\
\label{eqn::R11-W}  \Pr(R_i=(1,1) \mid A_i)&=& \Pr(Y_i(1)=1 \mid A_i).
\end{eqnarray}
Under Assumption~\ref{asm::ignorability}, we have
\begin{eqnarray}
\label{eqn::R00-W-identification}\Pr(Y_(d)=y \mid A_i)\ = \ \E \{\Pr(Y_i=y \mid D_i=d, \bX_i)\mid A_i\},
\end{eqnarray}
where we assume $\bX_i$ contains $A_i$.  Plugging
Equation~\eqref{eqn::R00-W-identification} into
Equations~\eqref{eqn::R00-W}~to~\eqref{eqn::R11-W} yields the formulas
in Theorem~\ref{thm::indep}. \QEDB


\section{Proof of Theorem~\ref{thm::learning}}
From the law of total probability, we have
\begin{eqnarray*}
\Pr(\delta(\bV_i)=1 \mid R_i=r, A_i)& =& \E \left\{ \Pr(D_i=1 \mid \bV_i,R_i=r, A_i) \mid R_i=r, A_i  \right\} \\
& =& \sum_{\bv} \Pr(\delta(\bV_i)=1 \mid \bV_i=\bv, A_i) \Pr(\bV_i=\bv \mid R_i=r, A_i)\\
&=&  \frac{\sum_x \Pr(\delta(\bV_i)=1 \mid \bV_i=\bv, A_i) \Pr(\bV_i=\bv \mid  A_i)\Pr(R_i=r\mid \bV_i=\bv,A_i)}{\Pr(R_i=r\mid   A_i)}\\
&=&  \frac{\sum_x \Pr(e_r(\bV_i,A_i)\delta(\bV_i) \mid \bV_i=\bv, A_i) \Pr(\bV_i=\bv \mid  A_i)}{\Pr(R_i=r\mid   A_i)}\\
&=&  \frac{\E(e_r(\{bV_i,A_i)\delta(\bV_i) \mid A_i\}}{\E\{e_r(\bV_i,A_i) \mid A_i \}}\\
&=& \E\left[ \frac{e_r(\bV_i,A_i)}{ \E\{e_r(\bV_i,A_i) \mid A_i \}}  \delta(\bV_i) \mid A_i \right],
\end{eqnarray*}
where can replace the summation with integral for continuous $\bV_i$. \QEDB

\end{document}